\documentclass[useAMS,usenatbib,usegraphicx]{mn2e}

\usepackage{times}

\newcommand\mujy{{$\umu$Jy}}
\newcommand\wtheta{{$w(\theta)$}}
\newcommand\aas{{A\&A}}
\newcommand\aass{{A\&AS}}
\newcommand\aj{{AJ}}
\newcommand\apj{{ApJ}}
\newcommand\apjl{{ApJL}}
\newcommand\apjs{{ApJS}}
\newcommand\mnras{{MNRAS}}
\newcommand\pasp{{PASP}}

\title[Galaxy clustering at 3.6~$\mu$m]{Clustering of galaxies at
  3.6 microns in the Spitzer Wide-area Infrared Extragalactic legacy
  survey}
\author[Waddington et al.]{I.~Waddington,$^1$
  S.~J.~Oliver,$^1$ T.~S.~R.~Babbedge,$^2$
  F.~Fang,$^{3}$ D.~Farrah,$^{4}$ 
\newauthor A.~Franceschini,$^{5}$ E.~A.~Gonzalez-Solares,$^{6}$ 
  C.~J.~Lonsdale,$^{7,8}$ G. Rodighiero,$^{5}$ 
\newauthor M.~Rowan-Robinson,$^{2}$ D.~L.~Shupe,$^{3}$
  J.~A.~Surace,$^{3}$ M.~Vaccari,$^5$ C.~K.~Xu$^{7}$\\
$^1$Astronomy Centre, University of Sussex, Brighton BN1 9QH, UK\\
$^2$Astrophysics Group, Blackett Laboratory, Imperial College London,
  Prince Consort Road, London SW7 2BW, UK\\
$^3$Spitzer Science Center, California Institute of
  Technology, MS 220-6, Pasadena, CA 91125, USA\\
$^4$Department of Astronomy, Cornell University, Space Sciences
  Building, Ithaca, NY 14853, USA\\
$^5$Dipartimento di Astronomia, Universita di Padova, Vicolo
  Osservatorio 5, I-35122 Padua, Italy\\
$^6$Institute of Astronomy, University of Cambridge, Madingley Road,
  Cambridge CB3 0HA, UK\\
$^7$Infrared Processing \& Analysis Center, California Institute of
  Technology, MS 100-22, Pasadena, CA 91125, USA\\
$^8$Center for Astrophysics \& Space Sciences, University of
  California San Diego, La Jolla, CA 92093--0424, USA\\
}

\begin{document}

\date{Accepted ---; Received ---; in original form ---.}

\pagerange{\pageref{firstpage}--\pageref{lastpage}}

\pubyear{2006}

\maketitle

\label{firstpage}

\begin{abstract}
We investigate the clustering of galaxies selected in the 3.6~\micron\
band of the {\it Spitzer\/} Wide-area Infrared Extragalactic (SWIRE)
legacy survey.  The angular two-point correlation function is
calculated for eleven samples with flux limits of $S_{3.6}\ge
4$--400~\mujy, over an 8~square degree field.  The angular clustering
strength is measured at $>5$-$\sigma$ significance at all flux limits,
with amplitudes of $A=(0.49$--$29)\times10^{-3}$ at one degree, for a
power-law model, $A\theta^{-0.8}$.  We estimate the redshift
distributions of the samples using phenomological models, simulations
and photometric redshifts, and so derive the spatial correlation
lengths.  We compare our results with the GalICS (Galaxies In
Cosmological Simulations) models of galaxy evolution and with
parameterized models of clustering evolution.  The GalICS simulations
are consistent with our angular correlation functions, but fail to
match the spatial clustering inferred from the phenomological models
or the photometric redshifts.  We find that the uncertainties in the
redshift distributions of our samples dominate the statistical errors
in our estimates of the spatial clustering.  At low redshifts (median
$z\le0.5$) the comoving correlation length is approximately constant,
$r_0=6.1\pm0.5h^{-1}$~Mpc, and then decreases with increasing redshift
to a value of $2.9\pm0.3h^{-1}$~Mpc for the faintest sample, for which
the median redshift is $z\sim1$.  We suggest that this trend can be
attributed to a decrease in the average galaxy and halo mass in the
fainter flux-limited samples, corresponding to changes in the relative
numbers of early- and late-type galaxies.  However, we cannot rule out
strong evolution of the correlation length over $0.5<z<1$.
\end{abstract}


\section{Introduction}

Galaxies are not distributed randomly across the sky.  At least at low
redshifts, they appear to trace distinct patterns: galaxy clusters are
connected to each other by long, filamentary structures of galaxies,
interspersed with large voids in which few or no galaxies are seen. A
plausible theoretical motivation has arisen for the formation of such
large-scale structures (LSS) of galaxies, namely that the galaxies are
tracing an underlying distribution of dark matter. In their most
modern form, models for the formation of these large scale structures
postulate that the evolution of the dark matter density field is
inextricably linked to the formation and evolution of the galaxies
themselves \citep[e.g.,][]{Cole00,Granato00,Hatton03}. These models
generally invoke some variation of the biased hierarchical paradigm,
in which overdensities, or `halos', in the dark matter distribution
undergo successive mergers over time to build halos of increasing
mass, with galaxies forming from the baryonic matter in these halos.

From these models, and recent observations, it is clear that the
relationship between the properties of galaxies, and the properties of
the dark matter halos in which they reside, is subtle, and is an area
in which observational constraints are particularly valuable in
constraining models. We would like to know, observationally, what sort
of galaxy occupies what sort of halo as a function of redshift, and
how the properties of galaxies change with both redshift and the
masses of their parent halos.  One method that has proven especially
useful in providing such observational constraints is measuring
clustering amplitudes. Fundamentally, the biased hierarchical paradigm
requires that overdensities of dark matter should themselves cluster
together on the sky, with the strength of clustering depending on
their mass \citep{Kaiser84,Bardeen86}. In principle, we can make great
strides in understanding the relationship between galaxies and the
underlying dark matter distribution by measuring the clustering
strength of galaxies selected in a particular way, and relating this
to theoretical predictions for halo clustering \citep{Benson01}.

Such observations are, however, not straightforward to perform.
Reliable clustering measurements require large, homogeneous samples of
sources selected over large enough areas of sky to sample a range of
dark matter density regimes. Ideally, we would like such observations
to be performed in the near- and mid-infrared: the (restframe)
near-infrard is most sensitive to evolved stars and so can pick up
large samples of passively evolving systems, whereas the mid-infrared
is sensitive to the dusty, active sources in which the stars and
central black holes in (at least some) passively evolving systems are
thought to form. Infrared observatories available up to now, however,
have not been capable of mapping large enough areas to the required
depths to find sufficient numbers of sources, or in enough bands to
even crudely discriminate between different populations.

The launch of the {\it Spitzer\/} Space Telescope \citep{Werner04}
offers the potential to overcome these problems, due to its ability to
map large areas of sky in the infrared to greater depths than any
previous observatory, and in multiple bands so that dusty, active
systems can be differentiated from passively evolving systems. The
{\it Spitzer\/} Wide-area Infrared Extragalactic (SWIRE) survey
\citep{Lonsdale03,Lonsdale04} is the largest of the {\it Spitzer\/}
Space Telescope's six Cycle~1 legacy programmes.  The survey covers a
total area of 49~square degrees, split between six fields, in all
seven of {\it Spitzer's\/} imaging bands (3.6--160~\micron). The area
and depth of SWIRE combine to produce a survey of significant comoving
volume, 0.2$h^{-3}$~Gpc$^3$ over $0<z<2$, and spatial scales of
$\sim100h^{-1}$~Mpc at $z\ge1$.

A principal goal of SWIRE is to study the clustering behaviour of a
variety of extragalactic populations.  In \citet{Oliver04} we
presented the first detection of galaxy clustering in the survey,
measuring a two-point angular correlation function at 3.6~\micron\
from our validation data, and in \citet{Farrah06} we presented results
on the clustering of Ultraluminous Infrared Galaxies (ULIRGs) at
$z>1$.  \citet{Fang04} presented angular correlation functions at
3.6--8.0~\micron\ from the 4-square degree {\it Spitzer\/} First Look
Survey.

In this paper we extend the analysis of the 3.6-\micron\ clustering to
larger scales and fainter flux limits (higher redshifts).  We begin
with a summary of definitions and formalisms in
section~\ref{sec:defs}, then in section~\ref{sec:sample} we discuss
the sample selection, including star/galaxy separation and the angular
selection function.  In section~\ref{sec:corrfunc} we present our
measurements of the two-point angular correlation function.  We
compare our results with previous measurements in the K-band
(section~\ref{sec:kband}) and with the GalICS semi-analytical
simulations (section~\ref{sec:galics}).  The angular clustering
amplitudes are used to estimate the spatial correlation lengths, which
are then compared with simple parameterized models of clustering
evolution, the GalICS simulations and results from the literature
(section~\ref{sec:r0}).  Section~\ref{sec:conclude} draws together
some conclusions from our analysis.  We use
$H_0=100h^{-1}$~km~s$^{-1}$~Mpc$^{-1}$ with $\Omega_M=0.3$ and
$\Omega_\Lambda=0.7$.  Magnitudes are in the AB system unless
otherwise noted.

\section{Definitions and Limber's equation}\label{sec:defs}

The spatial two-point correlation function $\xi(r,z)$ is defined
through the joint probablility
\begin{equation}
dP(r,z) = N^2 [ 1 + \xi(r,z) ] dV_1 dV_2
\label{eq:dpxi}
\end{equation}
of finding a galaxy in the volume element $dV_1$ and a second galaxy
in the volume element $dV_2$ separated by a distance $r$ at a redshift
$z$, where $N(z)$ is the mean number density of sources
\citep[e.g.,][]{Phillipps78}.  In comoving coordinates, the
correlation function can be parameterized as
\begin{equation}
\xi(r,z) = \left( r \over r_0 \right)^{-\gamma}
(1+z)^{\gamma-(3+\epsilon)}
\label{eq:xi}
\end{equation}
where $r_0$ measures the strength of the clustering at $z=0$, $\gamma$
measures the scale-dependence and $\epsilon$ parameterizes the
evolution with redshift \citep[e.g.,][]{Phillipps78,Overzier03}.

Several special values of $\epsilon$ have particular interpretations.
(1)~$\epsilon=0$ is the stable clustering model, where the correlation
function is fixed in proper coordinates and clustering grows stronger
as the background mass distribution expands with the universe.
(2)~$\epsilon=\gamma-3$ is the comoving case, where clustering remains
constant in comoving coordinates and simply expands with the universe.
(3)~$\epsilon=\gamma-1$ is the linear growth model, which corresponds
to the application of linear perturbation theory to a scale-free power
spectrum in an Einstein-de~Sitter universe.  We note that these models
are qualitative indicators of possible evolution scenarios, rather
than realistic clustering models \citep{Moscardini98}.

The angular two-point correlation function \wtheta\ is a measure of
the number of pairs of galaxies with separation $\theta$ compared with
that expected for a random distribution.  It is defined through the
joint probability
\begin{equation}
dP(\theta) = N_\Omega^2 [ 1 + w(\theta) ] d\Omega_1 d\Omega_2
\label{eq:dptheta}
\end{equation}
of finding a galaxy in solid angle $d\Omega_1$ and a second galaxy in
solid angle $d\Omega_2$ separated by an angle $\theta$, where
$N_\Omega$ is the mean number density of sources (per steradian) in
the survey \citep[e.g.,][]{Phillipps78}.  If \wtheta\ is zero, the
distribution of galaxies is unclustered.

The angular correlation function, \wtheta, is the projection along
the line of sight of the spatial correlation function, $\xi(r,z)$, and
can be calculated from Limber's equation \citep{Limber53,Phillipps78}.
If $\xi(r,z)$ is parameterized as a power-law, as above, then \wtheta\
is also a power-law
\begin{equation}
w(\theta) = A \theta^{1-\gamma}.
\label{eq:wtheta}
\end{equation}
The amplitude, $A$, of the angular correlation function can be
expressed \citep[following, e.g.,][]{Efstathiou91} as
\begin{equation}
A = { r_0^\gamma\ f \over c}\ { \int^\infty_0
  H(z)\ (1+z)^{-(2+\epsilon)}\ D_A^{1-\gamma}\ (dN/dz)^2\ dz\ \over \left[
  \int^\infty_0 (dN/dz)\ dz \right]^2 }
\label{eq:limber}
\end{equation}
where 
\begin{equation}
f = { \sqrt{\pi}\ \Gamma(
  [\gamma-1]/2 ) \over \Gamma(\gamma/2) }
\label{eq:limberfactor}
\end{equation}
with $\Gamma$ being the standard gamma function.  Here, $dN/dz$ is the
redshift distribution, $D_A$ is the angular diameter distance, and $H(z)$
is the Hubble parameter,
\begin{equation}
H(z) = H_0 \sqrt{ \Omega_M(1+z)^3 + \Omega_k(1+z)^2 + \Omega_\Lambda }
\label{eq:hubble}
\end{equation}
(where we have neglected the radiation energy density).  The spatial
correlation length, $r_0$, can thus be calculated from the amplitude
of the angular correlation function (eq.~\ref{eq:limber}) if the
redshift distribution, $dN/dz$, of the sources in the survey is known.

\section{Sample selection}\label{sec:sample}

\subsection{{\it Spitzer\/} observations}

The SWIRE-EN1 field has an area of $\sim$9~deg$^2$ and is coincident
with one of five fields observed as part of the European Large-Area
ISO (Infrared Space Observatory) Survey, ELAIS
\citep{Oliver00,Rowan-Robinson04}.  The nominal field centre is $\rm
16^h\,11^m\,00^s$ $+55^\circ\,00\arcmin\,00\arcsec$ (J2000).  The
field was mapped by {\it Spitzer\/} at 3.6, 4.5, 5.8 and 8.0~\micron\
with the Infrared Array Camera \citep[IRAC,][]{Fazio04} on 2004
January 14--20, and at 24, 70 and 160~\micron\ with the Multiband
Imaging Photometer \citep[MIPS,][]{Reike04} on 2004 January 21--28 and
2004 July 29.  The data can be retrieved from the {\it Spitzer\/}
archive with a Program Identification (PID) of 185, and the enhanced
data products (image mosaics and catalogues) are available from the
{\it Spitzer\/} Science
Center.\footnote{http://ssc.spitzer.caltech.edu/legacy}

Full details of the observations and data processing are given in
\citet{Surace06}; here we summarize the essential details.  The
SWIRE-EN1 field was mapped by IRAC with a large grid of pointings, and
at each grid point two 30-second images were taken, each one
consisting of multiple dithered exposures.  The entire grid was
repeated in two epochs, offset by half an array width.  Thus, for any
point on the sky there are a minimum of four independent sightings
(images), and these sightings occur on widely spaced parts of the
detector array in order to minimize instrumental signatures. The
entire survey has a minimum depth of four coverages, equal to
120~seconds of exposure time.  In some areas this can be as high as
sixteen coverages, or 480 seconds.

The IRAC data were reduced and flux-calibrated by the {\it Spitzer\/}
Science Center.  Further processing of the individual images removed a
number of effects (mostly due to bright objects, primarily stars) that
remained in the pipeline products \citep{Surace06}.  The images were
then coadded into sixteen large mosaics (or `tiles') of approximately
$0.8\times0.8$~square degrees each.  Sources were detected and their
photometry measured with the {\sc SExtractor} package
\citep{Bertin96}.  We used the flux measured in a circular aperture of
3\farcs8 diameter, unless the source was significantly extended, in
which case we used the flux within the Kron aperture.  The source
catalogue was a superset of the SWIRE Data Release 2 \citep{Surace05},
containing fainter objects than published at that time.

\subsection{Sample definition}\label{sec:fluxlimits}

We analyzed eleven flux-limited samples selected at 3.6~\micron\ from
the SWIRE-EN1 catalogue (Table~\ref{tab:amplitudes}).  The deepest
sample contained sources with flux densities $S_{3.6}\ge 4.0$~\mujy\
(or $m_{36}<22.4$~mag), this limit being defined by the flux density
at which the differential completeness is approximately 50 per cent.
At this level, the integral completeness is 82 per cent (see
Fig.~\ref{intcomplete} and discussion in section~\ref{sec:selection}).
Flux intervals of $\Delta\log S_{3.6}=0.2$ or $\Delta m_{36}=0.5$ were
used, corresponding to intervals of $\Delta z\sim0.1$ in the median
redshift of the samples (see section~\ref{sec:zdists}).


\begin{table}
\caption{Angular and spatial clustering strengths for 
  each of the samples.  $S_{36}$ are the flux limits, $A$ 
  are the amplitudes of the angular correlation functions 
  and $AC$ are the integral constraints. 
  $\left<z\right>$ are the median redshifts and $r_0$ are the 
  spatial correlation lengths, derived from the GalICS 
  redshift distributions for the $S_{36}=4.0$--15.9~\mujy\ samples 
  and from the 
  {\sc ImpZ} redshift distributions for the brighter samples 
  (see section~\ref{sec:r0}).}
\begin{tabular}{ccccc} \hline
$S_{36}$ & $A$         & $AC$ &  $\left<z\right>$ & $r_0$        \\ 
\mujy\   & $10^{-3}$ & $10^{-3}$ &                    & $h^{-1}$~Mpc \\ \hline
  4.0&$ 0.49\pm0.10$&$ 0.42\pm0.09$&1.00&$2.93\pm0.34$\\
  6.3&$ 0.63\pm0.14$&$ 0.54\pm0.12$&0.90&$3.18\pm0.38$\\
 10.0&$ 0.98\pm0.14$&$ 0.84\pm0.12$&0.80&$3.84\pm0.31$\\
 15.9&$ 1.48\pm0.08$&$ 1.26\pm0.07$&0.70&$4.48\pm0.14$\\
 25.2&$ 1.90\pm0.17$&$ 1.63\pm0.15$&0.65&$4.78\pm0.24$\\
 40.0&$ 2.47\pm0.35$&$ 2.12\pm0.30$&0.60&$5.28\pm0.41$\\
 63.4&$ 4.73\pm0.19$&$ 4.09\pm0.17$&0.50&$6.70\pm0.15$\\
100.5&$ 7.40\pm0.29$&$ 6.39\pm0.25$&0.38&$6.47\pm0.14$\\
159.2&$11.31\pm0.93$&$ 9.77\pm0.81$&0.28&$5.58\pm0.25$\\
252.4&$17.98\pm1.90$&$15.54\pm1.64$&0.20&$5.59\pm0.33$\\
400.0&$29.12\pm2.61$&$25.16\pm2.25$&0.17&$6.15\pm0.31$\\
  \hline
\end{tabular}
\label{tab:amplitudes}
\end{table}

The flux limit of the brightest sample was set by the need to have at
least 1000 sources in order to measure the ampitude of the angular
correlation function with more than 3-$\sigma$ significance.  This
limit on the minimum number of sources was determined empirically, by
calculating \wtheta\ for different subsets of the data, varying both
the total number of sources and the width of the angular bins in each
subset.  The brightest sample contained 1501 sources with $S_{3.6}\ge
400.0$~\mujy\ (or $m_{36}<17.4$~mag).

\subsection{Angular selection function}\label{sec:selection}

In order to calculate clustering statistics, we require an angular
selection function that would describe the distribution of sources in
the survey if there was no clustering.  This takes the form of a mask,
where the value of the mask at each position is the relative
probability of finding a source at that location on the sky in the
absence of clustering.  This mask is then used to simulate a random
(i.e., unclustered) catalogue of objects (section~\ref{sec:ls}).  The
probability of detecting a source at any given position depends on the
completeness of the survey at that point, which is a function of the
noise in the image.  A noise map was calculated from the coverage map
(i.e.\ a map of the integration time) and the completeness function
was found from simulations, as follows.

Coverage maps were generated for each individual mosaic, recording the
number of independent images contributing to each pixel, after taking
into account the complex dithering pattern and any missing data due to
cosmic ray rejection.  The combined map for the whole SWIRE-EN1 field
was rebinned by a factor of 5 to a pixel scale of 3~arcsec, reducing
the size of the map so that it would fit into memory.  This rebinning
effectively smoothed the coverage map on a scale of 3~arcsec, closely
matching the 3\farcs8-diameter photometric aperture.  Pixels with mean
coverage less than 2.95 were excluded from the mask (this allowed for
up to one of the images in the full coverage areas to be flagged and
rejected due to a cosmic ray).  The noise in the mosaics scales with
the integration time, $t$, as $\sigma\propto1/\sqrt{t}$ and so varies
with coverage (number of codded images), $\kappa$, also as
$\sigma\propto1/\sqrt{\kappa}$ \citep{Surace05}.  The coverage mask
was then used as a proxy for a noise map.

\begin{figure}
\includegraphics[width=\hsize]{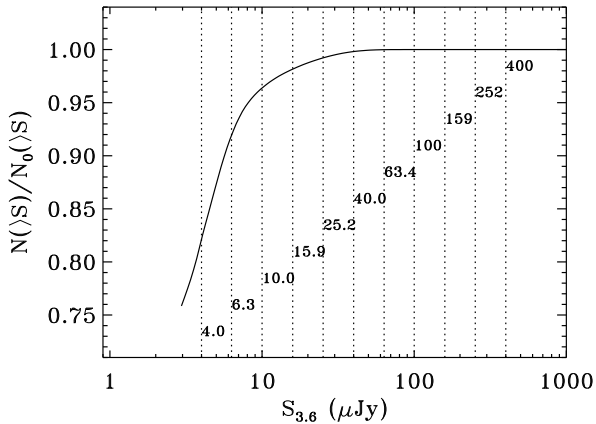}
\caption{The integral completeness function at 3.6~\micron.  $N(>S)$
  is the cumulative number of sources with fluxes $\ge S_{3.6}$, and
  $N_0(>S)$ is similarly the cumulative number of sources after
  correction for incompleteness.  The dotted lines show the
  flux limits of our samples.}
\label{intcomplete}
\end{figure}

We calculated the completeness function by simulating artificial
sources and adding them into the SWIRE images.  The source extraction
stage of the analysis was then repeated, and the new source list was
compared with the known positions and fluxes of the artificial
sources.  The fraction of simulated sources that were recovered by the
source extraction was computed as a function of flux and coverage
(noise).  Fig.~\ref{intcomplete} shows the integral completeness as
a function of flux limit, for an average coverage of 5.0 pointings.
The survey is 99\% complete at 22~\mujy, 95\% complete at 8.1~\mujy,
and the integral completeness falls to 82\% for our faintest sample at
4.0~\mujy.  The simulation results also confirmed that the
completeness, $f$, scaled as expected with coverage, $\kappa$, and
flux limit, $S$, as $f(S,\kappa)=f(\sqrt{\kappa}S,1)$.  At every point
in the coverage map, we calculated the integral completeness given the
flux limit and coverage.  This was then the relative probability that
a source in the survey could have been found at that location, in the
absence of clustering.

The final contribution to the angular selection function was to
exclude circular regions around bright point sources.  For this we
used the Two-Micron All Sky Survey \citep[2MASS,][]{Skrutskie06}, masking
a circle around all $K\le12$ point sources within a radius, $R$, given
by $\log R({\rm arcsec}) = 3.1 - 0.16 K$.  This radius was determined
by visual inspection of the SWIRE images to find the distance at which
the star's PSF merges into the background.  Note that these values are
about a factor of two larger than that used in the public SWIRE
catalogue \citep{Surace05}, giving a more conservative mask.

\subsection{Star/galaxy separation}

The presence of stars in the source catalogue artificially dilutes the
strength of the galaxy correlations, so it was necessary to remove the
stars before performing the clustering analysis.  To make the best use
of the large survey area, we did not want to be restricted to the
limited area with optical coverage, so we developed a procedure to
remove stars using only infrared criteria.  Our goal was to minimize
the number of stars remaining in the sample, but without removing an
excessive number of galaxies.

We explored a range of magnitude, colour and stellarity selection
criteria in order to identify stars in the infrared data, developing a
three-stage process to remove them.  First, the SWIRE catalogue was
cross-correlated with the 2MASS survey to classify the bright sources.
Objects flagged as being extended in 2MASS are galaxies and bright
($K\le14$~mag) point sources are stars.  Second, faint 2MASS sources
were classified based on their near- to mid-infrared colours and
stellarity at 3.6~\micron.  Point-like (${\rm stellarity}>0.94$)
sources are identified as stars.  Sources that are both blue
($J-m_{36} < -1.50$ or $H-m_{36} < -2.2$) and not clearly resolved
(${\rm stellarity}>0.06$) are also stars.  Third, stars fainter than
2MASS were classified based on their mid-infrared colours and
3.6~\micron\ stellarity, where blue compact sources are stars.  We
used three flux bins ($m_{45}\le19.5$, $19.5<m_{45}\le20.0$ and
$20.0<m_{45}\le23.0$) with colour cuts of $m_{36}-m_{45}<-0.35$,
$-0.30$ and $-0.25$, and ${\rm stellarity}>0.8$, 0.8 and 0.7
respectively to identify stars.

A subset of the SWIRE-EN1 field has optical imaging data which we used
to estimate the effectiveness of our infrared star/galaxy separation.
The Isaac Newton Telescope Wide Angle Survey
\citep{McMahon01,Gonzalez-Solares04b} observed 6.4~deg$^2$ of the
SWIRE-EN1 survey in five optical bands
($U,g^\prime,r^\prime,i^\prime,Z$) to $r^\prime\simeq23.5$ mag.  We
identified optical counterparts to the SWIRE sources and selected
those sources with high optical ($i$-band) stellarity according to the
{\sc SExtractor} source extraction software \citep{Bertin96}.  These
sources constituted a reference list of stars, against which we
compared the infrared classifications.  At bright fluxes ($m_{36}<21$)
the stellar contamination in the galaxy sample was estimated to be
$<3$\%, rising to 4\% at $m_{36}=22$--23.  Fainter than $m_{36}=23$,
the total star counts are $\la3$\% of the galaxy counts
\citep{Fazio04} so the contamination is still low, even though we can
no longer identify stars at these faint magnitudes.  We also compared
our galaxy sample with a list of stars identified by the {\sc ImpZ}
photometric redshift estimation code
\citep{Babbedge04,Rowan-Robinson05}.  Again we found that the stellar
contamination in our infrared galaxy catalogue was only 2--4\% at
$m_{36}<23$~mag.

At all fluxes, approximately 10\% of the galaxies were rejected by our
star selection criteria.  There was a slight bias towards rejecting
blue (in near- to mid-infrared colours) compact galaxies, but the
fraction of galaxies rejected was sufficiently small that this is not
expected to significantly bias the measurement of the angular
correlation function.

\section{Angular correlation functions}\label{sec:corrfunc}

\subsection{Method}\label{sec:ls}

The angular correlation function (Eq.~\ref{eq:wtheta}) was estimated
by comparing the distribution of galaxies in the survey with
catalogues of random sources.  The random catalogues were simulated
using the angular selection function (Section~\ref{sec:selection}),
such that the angular distribution of the random sources reflected the
geometry and variable depth of the actual survey.  We compared each
real dataset with 1000 random catalogues, each containing the same
number of sources as the galaxy catalogue, ensuring that the
uncertainty in the correlation function was not dominated by the
scatter between the random catalogues.

A number of methods have been proposed to calculate the angular
correlation function; here we use the \citet{Landy93} estimator,
\begin{equation}
\hat{w}(\theta) = { DD - 2DR + RR \over RR }
\label{eq:ls}
\end{equation}
where $DD$ is the number of galaxy--galaxy pairs at separation
$\theta$, normalized by the total number of pairs over all
separations; and $DR$ and $RR$ are similarly the normalized number of
galaxy--random and random--random pairs respectively.

The computationally-intensive step in the calculation is counting the
number of pairs of sources, $DD$, $DR$ and $RR$.  For this we used the
{\sc npt} pair-counting code of \citet{Gray04} which uses kd-trees to
greatly accelerate the speed of the calculation compared with naive
counting methods.  We used twelve logarithmically-spaced bins in
$\theta$, over the range $0.001<\theta<4.0$~degrees, with
$\Delta\log\theta=0.3$.

\subsection{Error estimates and parameter fitting}

The large number of sources in the SWIRE dataset allowed us to
calculate the errors on the correlation function by comparing subsets
of the data.  For each of the six faint flux-limited samples
(4.0--40.0~\mujy) we divided the data into 9--25 subsamples of
10,000--15,000 sources each.  Two sampling methods were used.  First,
we randomly selected galaxies across the full field; this gave good
statistics on the larger scales ($>0.05$~deg).  Second, we obtained
better statistics on smaller scales by dividing each flux-limited
sample into a grid of 9--25 smaller regions and calculating \wtheta\
using all the sources within a sub-region.

For each sub-sampling method, this produced $n=9$--25 independent
estimates of the correlation function, $\hat{w}_i(\theta_j)$, where $i$
labels the subsample and $j$ labels the angular bin.  From these $n$
estimates, we calculated the mean $\bar{w}(\theta_j) =
\sum_{i=1}^n \hat{w}_i(\theta_j) / n$ for each bin.  The covariance
between angular bins $\theta_j$ and $\theta_k$ is given by
\begin{equation}
\sigma_{jk}^2 = { 1 \over n - 1 } \sum_{i=1}^n [ \hat{w}_i(\theta_j)
- \bar{w}(\theta_j) ] [ \hat{w}_i(\theta_k) - \bar{w}(\theta_k) ].
\label{eq:cov1}
\end{equation}
As other authors have also found
\citep[e.g.,][]{Zehavi02}, the off-diagonal terms become increasingly
noisy for elements farther away from the diagonal, making inversion of
the covariance matrix unstable, so we only retained the covariances
between adjacent bins, i.e.\ we set $\sigma_{jk}^2 = 0$ for $|j-k|>1$.

The brighter flux-limited samples (63--400~\mujy) contained too few
sources to divide into independent samples, so to calculate the errors
we used the jackknife technique.  Each sample was divided into a grid
of $4\times4$ sub-areas and we calculated $\hat{w}_i(\theta_j)$
sixteen times, each time excluding a different sub-area.  The best
estimate of the correlation function is the mean of the
$\hat{w}_i(\theta_j)$ and the covariance is
\begin{equation}
\sigma_{jk}^2 = { n - 1 \over n } \sum_{i=1}^n [ \hat{w}_i(\theta_j)
- \bar{w}(\theta_j) ] [ \hat{w}_i(\theta_k) - \bar{w}(\theta_k) ]
\label{eq:cov2}
\end{equation}
\citep{Scranton02}.  Again, we only retained the covariances between
adjacent bins.

\begin{figure*}
\begin{minipage}{180mm}
\includegraphics[width=\hsize]{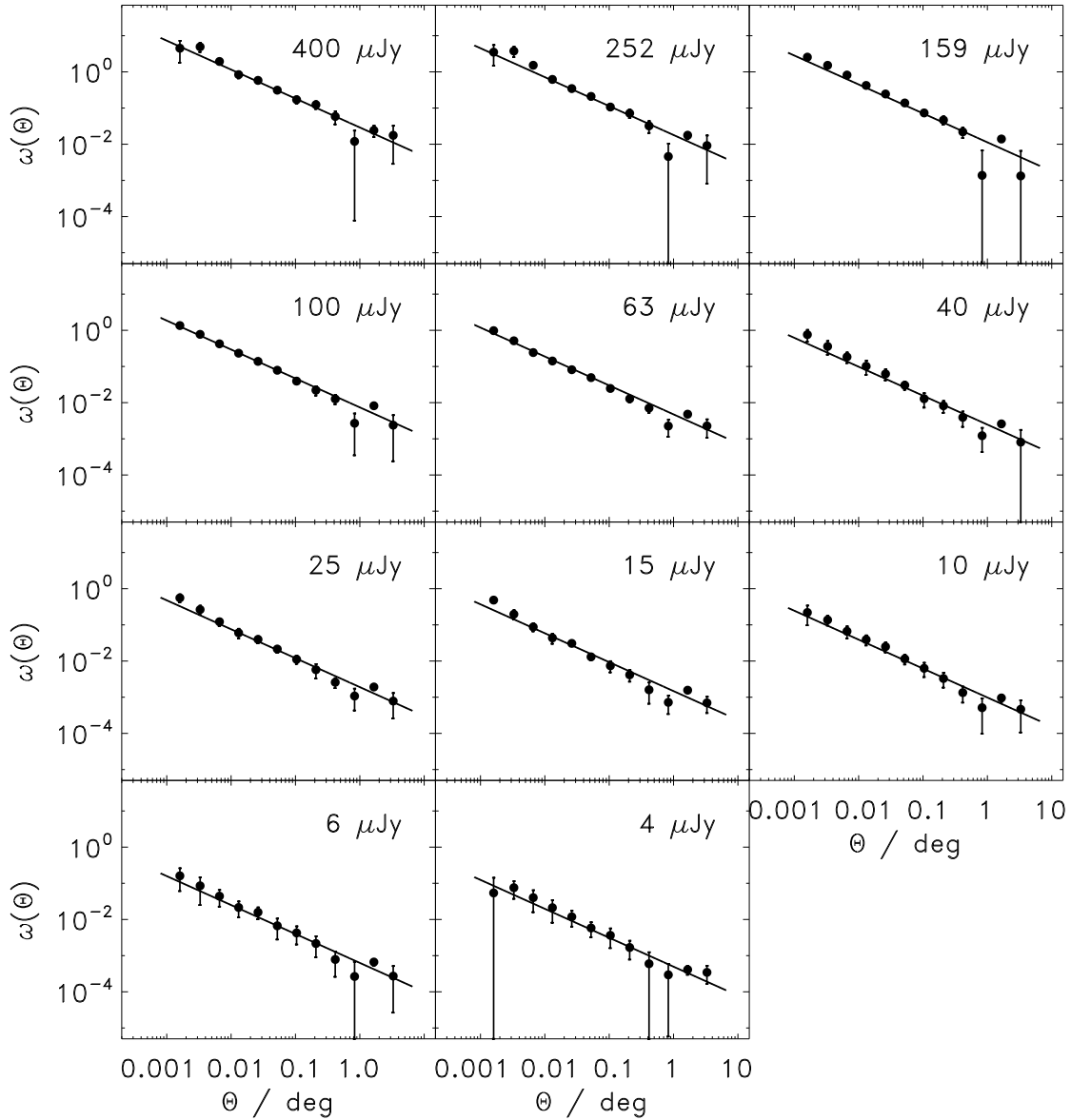}
\caption{Angular correlation functions \wtheta\ for the eleven samples
  with flux density limits of 4--400~\mujy.  Solid lines are the
  best-fitting power-law model with $\gamma=1.8$, and the data have
  been corrected for the integral constraint.}
\label{fig:wtheta}
\end{minipage}
\end{figure*}

In Fig.~\ref{fig:wtheta} we plot the angular correlation functions
for the eleven samples, corrected by the integral constraint discussed
below.  For the faint samples, we show the data from the full-field
subsampling, which give the best results on large scales and are
consistent on smaller scales with the results of the sub-region
sampling.

The angular correlation function is parameterized as a power law $A
\theta^{1-\gamma}$ (Eq.~\ref{eq:wtheta}), where the amplitude, $A$,
measures the strength of the clustering, the index, $\gamma$, measures
its scale-dependence, and $\theta$ is measured in degrees.  Due to the
finite size of the survey, the observed $\bar{w}(\theta)$ is a biased
estimator of the real correlation function, and can be modeled as
\begin{equation}
w_{\rm m}(\theta) = A ( \theta^{1-\gamma} - C ).
\label{eq:wthetacorr}
\end{equation}
The integral constraint, $AC$, can be estimated by doubly integrating
the (assumed) true $w_{\rm m}(\theta)$ over the area of the survey, a
calculation that can be done numerically using the random--random pair
counts,
\begin{equation}
C = { \sum_j N_{rr}(\theta_j) \theta_j^{1-\gamma} \over \sum_j
  N_{rr}(\theta_j) }
\label{eq:intcorr}
\end{equation}
where $N_{rr}(\theta_j)$ are the unnormalized counts and the summation
is over all the angular bins \citep{Roche99}.

We fitted the model (Eq.~\ref{eq:wthetacorr}) to the observed
correlation function $\bar{w}(\theta_j)$ for each of the data
samples by minimizing the generalized $\chi^2$, defined as
\begin{equation}
\chi^2 = \sum_{j=1}^n \sum_{k=1}^n [ w_{\rm m}(\theta_j) -
\bar{w}(\theta_j) ] H_{jk} [ w_{\rm m}(\theta_k) - \bar{w}(\theta_k) ]
\label{eq:chisq}
\end{equation}
where $H_{jk}$ is the inverse of the covariance matrix,
$(\sigma_{jk}^2)^{-1}$ \citep{Pollo05}.  A wide range of optical and
infrared surveys indicate that $\gamma=1.8$ and we found that this was
consistent with the present data, so fixed $\gamma$ to this value for
comparison with other surveys.  The best-fitting models are shown with
the data in Fig.~\ref{fig:wtheta} and the amplitudes and integral
constraints are listed in Table~\ref{tab:amplitudes}.


\begin{table*}
\begin{minipage}{115mm}
\caption{Spatial correlation lengths, $r_0$ ($h^{-1}$~Mpc), 
  and median redshifts, $\left<z\right>$, derived from the 
  SWIRE clustering amplitudes and each 
  of the redshift distributions.
  $S_{36}$ are the flux limits in \mujy.}
\begin{tabular}{ccccccccc} \hline
 &   \multicolumn{2}{c}{Xu et al.}           &
 \multicolumn{2}{c}{Franceschini et al.} &
 \multicolumn{2}{c}{GalICS}              &   \multicolumn{2}{c}{{\sc ImpZ}}                \\
 $S_{36}$        &   $\left<z\right>$ & $r_0$   &  $\left<z\right>$ & $r_0$   &  $\left<z\right>$ & $r_0$   &  $\left<z\right>$ & $r_0$   \\ \hline
  4.0&0.88&$2.85\pm0.33$&0.94&$2.95\pm0.34$&1.00&$2.93\pm0.34$&0.68&$2.46\pm0.28$\\
  6.3&0.84&$3.18\pm0.38$&0.88&$3.22\pm0.38$&0.90&$3.18\pm0.38$&0.68&$2.79\pm0.33$\\
 10.0&0.78&$3.97\pm0.32$&0.82&$3.92\pm0.32$&0.80&$3.84\pm0.31$&0.68&$3.48\pm0.28$\\
 15.9&0.74&$4.82\pm0.15$&0.74&$4.66\pm0.14$&0.70&$4.48\pm0.14$&0.65&$4.25\pm0.13$\\
 25.2&0.70&$5.32\pm0.27$&0.68&$5.09\pm0.26$&0.60&$4.72\pm0.24$&0.65&$4.78\pm0.24$\\
 40.0&0.66&$5.91\pm0.46$&0.62&$5.55\pm0.43$&0.50&$4.82\pm0.37$&0.60&$5.28\pm0.41$\\
 63.4&0.58&$7.70\pm0.17$&0.56&$7.33\pm0.17$&0.40&$5.78\pm0.13$&0.50&$6.70\pm0.15$\\
100.5&0.46&$8.52\pm0.19$&0.48&$8.27\pm0.18$&0.30&$5.87\pm0.13$&0.38&$6.47\pm0.14$\\
159.2&0.36&$8.75\pm0.40$&0.36&$8.49\pm0.39$&0.25&$5.66\pm0.26$&0.28&$5.58\pm0.25$\\
252.4&0.26&$8.58\pm0.50$&0.24&$7.97\pm0.47$&0.20&$5.61\pm0.33$&0.20&$5.59\pm0.33$\\
400.0&0.20&$8.51\pm0.42$&0.18&$7.16\pm0.36$&0.15&$5.71\pm0.28$&0.17&$6.15\pm0.31$\\
  \hline
\end{tabular}
\label{tab:amplitudes2}
\end{minipage}
\end{table*}

\section{Comparison with K-band surveys}\label{sec:kband}

\begin{figure}
\includegraphics[width=\hsize]{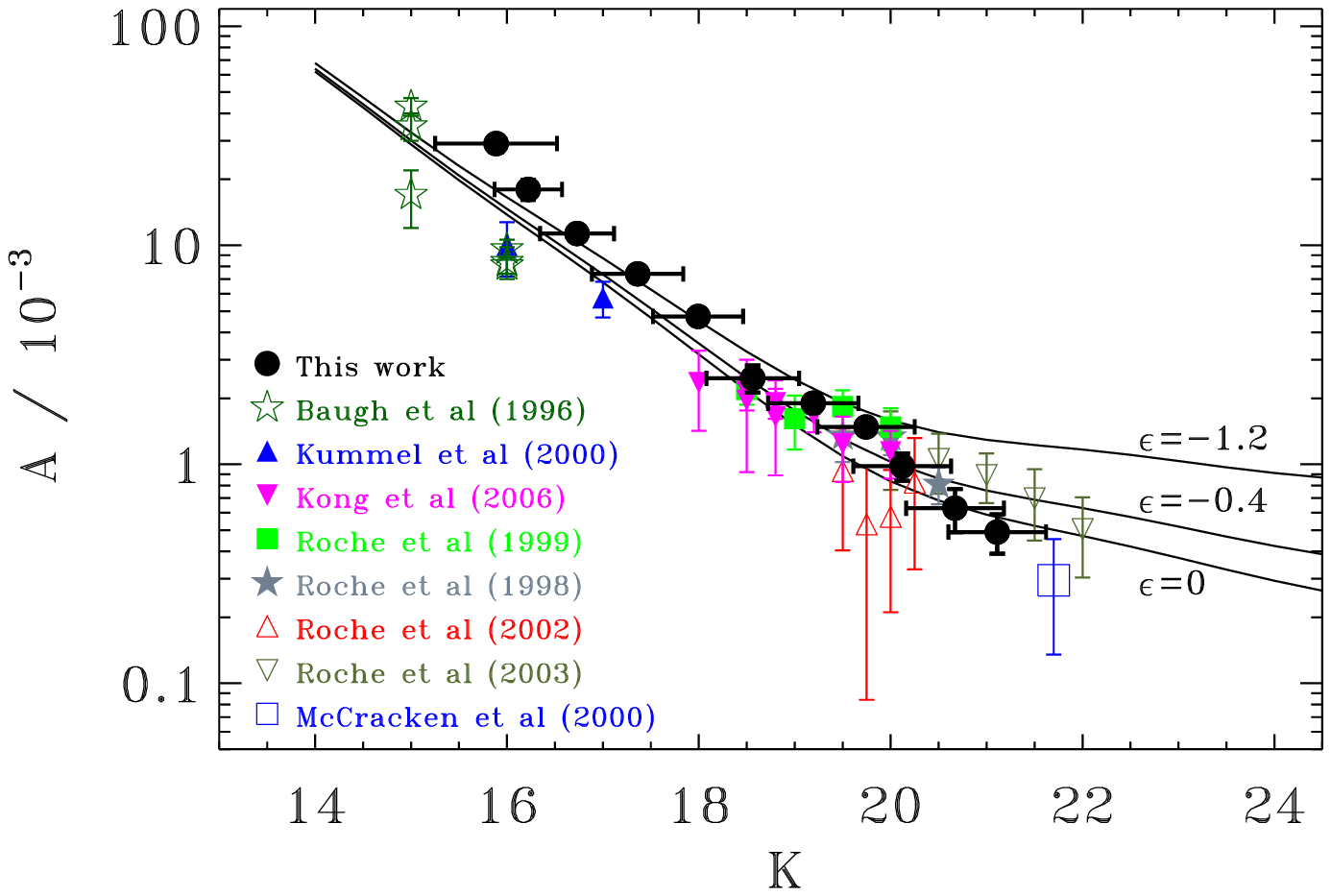}
\caption{The amplitude, $A$, of the angular correlation function, as a
  function of $K$-band limiting magnitude (Vega system), for our SWIRE
  data (solid circles) and surveys from the literature
  \citep{Baugh96,Kong06,Kummel00,McCracken00,Roche98,Roche99,Roche02,
  Roche03}.  The $K$-band limits of the SWIRE data have been estimated
  from average $K-m_{36}$ colours (section~\ref{sec:kband}).  The
  lines are models of clustering evolution from \citet{Roche03}.}
\label{fig:roche}
\end{figure}

The angular correlation function is the projection along the line of
sight of the spatial correlation function, $\xi(r,z)$, and is
dependent on both the redshift distribution and luminosity function of
galaxies in the survey and on the evolution of the spatial clustering.
At fainter flux limits the survey probes to higher redshifts (larger
volumes) and lower luminosities, and both these reduce the strength of
the projected clustering.  This is shown in Fig.~\ref{fig:roche} where
we plot the amplitude, $A$, of the angular correlation function
against the limiting magnitude for a range of $K$-band surveys.  We
compare the 3.6-\micron\ SWIRE results with previous $K$-band data,
due to the abundance of clustering measurements in $K$ and the
relatively few measurements at 3.6-\micron\ \citep{Fang04,Oliver04}.
The emission in both the $K$ and 3.6-\micron\ bands arises from the
old stars in a galaxy -- both bands are relatively insensitive to the
current star formation rate and are good tracers of the stellar mass.

We estimated the equivalent $K$-band limit (Vega system) for each of
our 3.6~\micron\ selected samples using the average $K-m_{36}$ colour
of SWIRE galaxies detected in the Early Data Release of the UKIRT
Infrared Deep Sky Survey \citep[UKIDSS;][]{Dye06}.  We overplot the
SWIRE clustering amplitudes on Fig.~\ref{fig:roche}, where the errors
in $K$-band magnitude correspond to the standard deviation of the
$K-m_{36}$ colour distributions.  Our data are consistent with these
other surveys, confirming that both bands are selecting similar galaxy
populations, with our new results having significantly smaller errors
in the amplitudes, due to the much larger survey area at fainter
fluxes.  The larger area of future UKIDSS data releases will enable us
to reduce the uncertainty in the equivalent $K$-band limits of SWIRE,
particularly at the bright end where there are relatively few sources
in UKIDSS at present.

Also plotted in Fig.~\ref{fig:roche} are three models of clustering
evolution (Eq.~\ref{eq:xi}), with $\epsilon=0$, $-0.4$ \& $-1.2$, from
figure~7 of \citet{Roche03}.  Following Eq.~\ref{eq:limber}, the
amplitude of the angular correlation function can be calculated if the
redshift distribution ($dN/dz$) of the survey is known.
\citet{Roche02,Roche03} predict the redshift distributions using a
simple galaxy evolution model, where they evolve a $K$-band galaxy
luminosity function according to stellar population synthesis codes,
given a star formation history and a galaxy merger rate.  We see that
the model that best fits the SWIRE data is their stable clustering
model ($\epsilon=0$), with their comoving model ($\epsilon=-1.2$)
rejected at $>5$-$\sigma$ at the faintest magnitudes ($K\sim21$).  We
note that these models of the angular clustering are dependent on a
range of parameters, not just the evolution of the spatial clustering,
$\epsilon$, so it is the \citet{Roche03} merger model with comoving
evolution that is a poor fit to the data, and this does not imply that
comoving clustering in general can be rejected.

\section{Comparison with GalICS mock catalogues}\label{sec:galics}

GalICS (Galaxies In Cosmological Simulations) is a hybrid model of
galaxy evolution which combines high-resolution N-body simulations of
the dark matter content of the universe with semi-analytic
prescriptions to describe the fate of the baryons within the dark
matter halos \citep{Hatton03}.  The simulations have 256$^3$ particles
of mass $8\times10^9$~M$_\odot$, in a 100$h^{-1}$~Mpc box with a
spatial resolution of 20$h^{-1}$~kpc.  Within each halo, some fraction
of the gas mass is cooled and turned into stars which then evolve.
The spectral energy distributions of these model galaxies are computed
by summing the contribution of all the stars they contain, tracking
their age and metallicity.  A mock catalogue is generated by
projecting a cone through the simulation at a series of timesteps
(redshifts), and calculating the properties of the galaxies `observed'
in the cone.  The GalICS project have made
available\footnote{http://galics.cosmologie.fr/} these 1-deg$^2$
cones, from which we have extracted mock catalogues of the SWIRE
survey.


\begin{table}
\caption{Angular and spatial clustering strengths for the 
  GalICS simulations.  $S_{36}$ are the flux limits, $A$ 
  are the amplitudes of the angular correlation functions, 
  $AC$ are the integral constraints, 
  $\left<z\right>$ are the median redshifts and $r_0$ are the 
  spatial correlation lengths.}
\begin{tabular}{ccccc} \hline
$S_{36}$ & $A$       & $AC$ &  $\left<z\right>$ & $r_0$        \\ 
\mujy\   & $10^{-3}$ & $10^{-3}$ &                    & $h^{-1}$~Mpc \\ \hline
  4.0&$ 0.77\pm0.03$&$ 1.56\pm0.06$&1.00&$3.76\pm0.08$\\
  6.3&$ 0.87\pm0.05$&$ 1.76\pm0.09$&0.90&$3.81\pm0.11$\\
 10.0&$ 1.28\pm0.06$&$ 2.57\pm0.12$&0.80&$4.44\pm0.11$\\
 15.9&$ 1.50\pm0.13$&$ 3.03\pm0.25$&0.70&$4.53\pm0.21$\\
 25.2&$ 1.87\pm0.13$&$ 3.77\pm0.25$&0.60&$4.67\pm0.17$\\
 40.0&$ 2.58\pm0.14$&$ 5.21\pm0.27$&0.50&$4.94\pm0.14$\\
 63.4&$ 3.85\pm0.42$&$ 7.76\pm0.86$&0.40&$5.15\pm0.32$\\
100.5&$ 4.18\pm0.53$&$ 8.43\pm1.07$&0.30&$4.27\pm0.30$\\
159.2&$10.67\pm2.07$&$21.52\pm4.18$&0.25&$5.48\pm0.59$\\
252.4&$13.47\pm2.42$&$27.16\pm4.88$&0.20&$4.78\pm0.48$\\
400.0&$15.92\pm4.06$&$32.07\pm8.18$&0.15&$4.08\pm0.58$\\
  \hline
\end{tabular}
\label{tab:galics}
\end{table}

We calculated the two-point angular correlation functions for the
eight GalICS catalogues, each time using the eleven flux-limited
samples corresponding to the flux limits listed in
Table~\ref{tab:amplitudes}.  We used the same method as for the SWIRE
data (section~\ref{sec:corrfunc}) and calculated errors and
covariances from the eight independent samples.  The correlation
functions were fitted by a power-law model (Eq.~\ref{eq:wthetacorr})
with fixed $\gamma=1.8$, to determine the amplitudes and the integral
constraints.  The results are given in Table~\ref{tab:galics}.

\begin{figure*}
\begin{minipage}{180mm}
\includegraphics[width=\hsize]{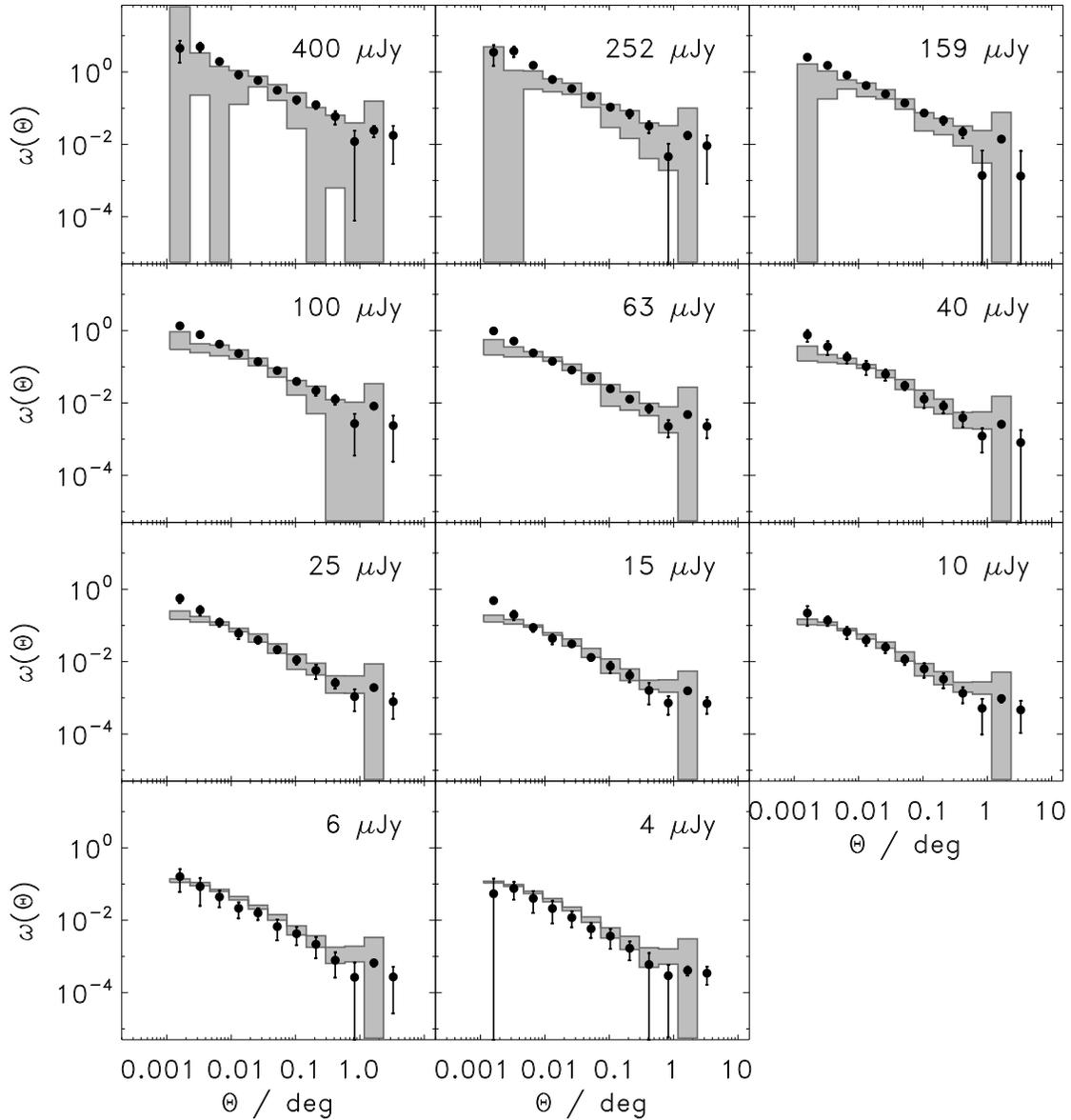}
\caption{Angular correlation functions \wtheta\ for the eleven
  flux-limited samples compared with the GalICS simulations.  The
  shaded regions are the one-sigma error bounds on \wtheta\ from
  the mock catalogues; the data points correspond to
  Fig.~\ref{fig:wtheta}.  Both the data and the simulation results
  have been corrected for their integral constraints.}
\label{fig:galics}
\end{minipage}
\end{figure*}

In Fig.~\ref{fig:galics} we compare the SWIRE correlation functions
(data points) with the GalICS results, where the shaded regions are
the one-sigma error bounds from the mock catalogues, and all datasets
have been corrected for their integral constraints.  At the brightest
flux limits ($S_{3.6}\ge 159$~$\mu$Jy), the large uncertainties in the
model correlation functions, particularly on small scales, are due to
the small size (1~sq.~deg.) of the simulations compared with the data
(8.1~sq.~deg.).  There are less than 1000 sources in each of these
bright samples, and it is seen that this is insufficient to measure
\wtheta\ accurately at $\theta\la 0.01$~deg.  Similarly, the greater
uncertainties in the models on large scales are due to the smaller
angular size of the GalICS catalogues.  The largest angular scale of
the simulations is 1.4~deg compared with 4.1~deg for the data,
although the measurement of \wtheta\ becomes uncertain on scales
much smaller than the maximum extent of either survey.

We compared the GalICS correlation functions with the SWIRE results
using a $\chi^2$ test.  If the two smallest scale bins are excluded
from the comparison, then the data and simulations do not differ
significantly.  Even for the faintest two samples (4 and 6~\mujy),
where the simulations lie consistently above the data in
Fig.~\ref{fig:galics}, the difference is not statistically
significant.  However the small-scale discrepencies between the
simulations and the data are worth noting.  The GalICS correlation
functions deviate from a power law at scales $\la 30$~arcsec at all
flux limits for which there is good data on small scales.  The
$\chi^2$ test gives the probability that the data and simulations are
drawn from the same distribution as only $10^{-2}$--$10^{-4}$ for
these small scales, for almost all samples brighter than $S_{36}\ge
15$~\mujy.

This lack of close ($\la100h^{-1}$~kpc) pairs in the GalICS
simulations, corresponding to the scale of galaxy groups and smaller,
was also observed by \citet{Blaizot06} in their comparison of the
GalICS angular correlation functions with those of the Sloan Digital
Sky Survey \citep{York00}.  This can be explained within the context
of Halo Occupation Distribution models \citep[e.g.,][]{Berlind02}.
Such models have shown that the galaxy correlation function can be
decomposed into two terms: (i) correlations between galaxies in
different halos (large scales), and (ii) correlations between pairs of
galaxies located within the same halo (small scales).
\citet{Blaizot06} showed that the GalICS simulations underestimate
this clustering of galaxies within a single halo, and this leads to
the turnover of the GalICS correlation function at small scales, as we
see in Fig.~\ref{fig:galics}.

There is a second factor which may contibute to the underestimate of
small-scale clustering in the simulation results.  The galaxy mass
resolution limit of GalICS corresponds to a limiting galaxy luminosity
($M_K<-22.7$ in the restframe $K$-band) and an absence of
low-mass/low-luminosity galaxies in the simulations would also reduce
the number of close pairs.  A luminous galaxy is far more likely to
have a low-luminosity companion, simply because their number density
is that much greater, and if these low-luminosity galaxies are missing
from the simulation, then there will be an absence of close pairs.  As
discussed below (Section~\ref{sec:r0}), the fainter samples contain
proportionately more low-luminosity galaxies than the bright samples,
and if these are missing from the GalICS models due to the mass
resolution limit, then that could also explain the trend for the
models to over-predict the clustering at 4 and 6~\mujy.

Overall, the GalICS simulations are in excellent agreement with the
angular clustering of 3.6-\micron\ selected galaxies in SWIRE.
However, the failure to resolve low-mass sources and to correctly
predict the clustering within a given halo, indicate that these data
are pushing the GalICS models to their limit.

\section{Spatial clustering}\label{sec:r0}

\subsection{Redshift distributions}\label{sec:zdists}

The angular correlation function is the projection along the line of
sight of the spatial correlation function.  We have used the inverse
of Limber's equation (Eq.~\ref{eq:limber}) to estimate the strength of
the spatial clustering, expressed as the correlation length,
$r_0$, from our measurements of the angular clustering amplitude, $A$.
Limber's equation is expressed in terms of the redshift distribution
of the sample, $dN/dz$, which incorporates the radial selection
function, i.e.\ the probability that a source at a given redshift
could have been detected in the survey.  We have used four independent
estimates of the redshift distribution to calculate the spatial
correlation lengths for each SWIRE sample.

\begin{figure*}
\begin{minipage}{180mm}
\includegraphics[width=\hsize]{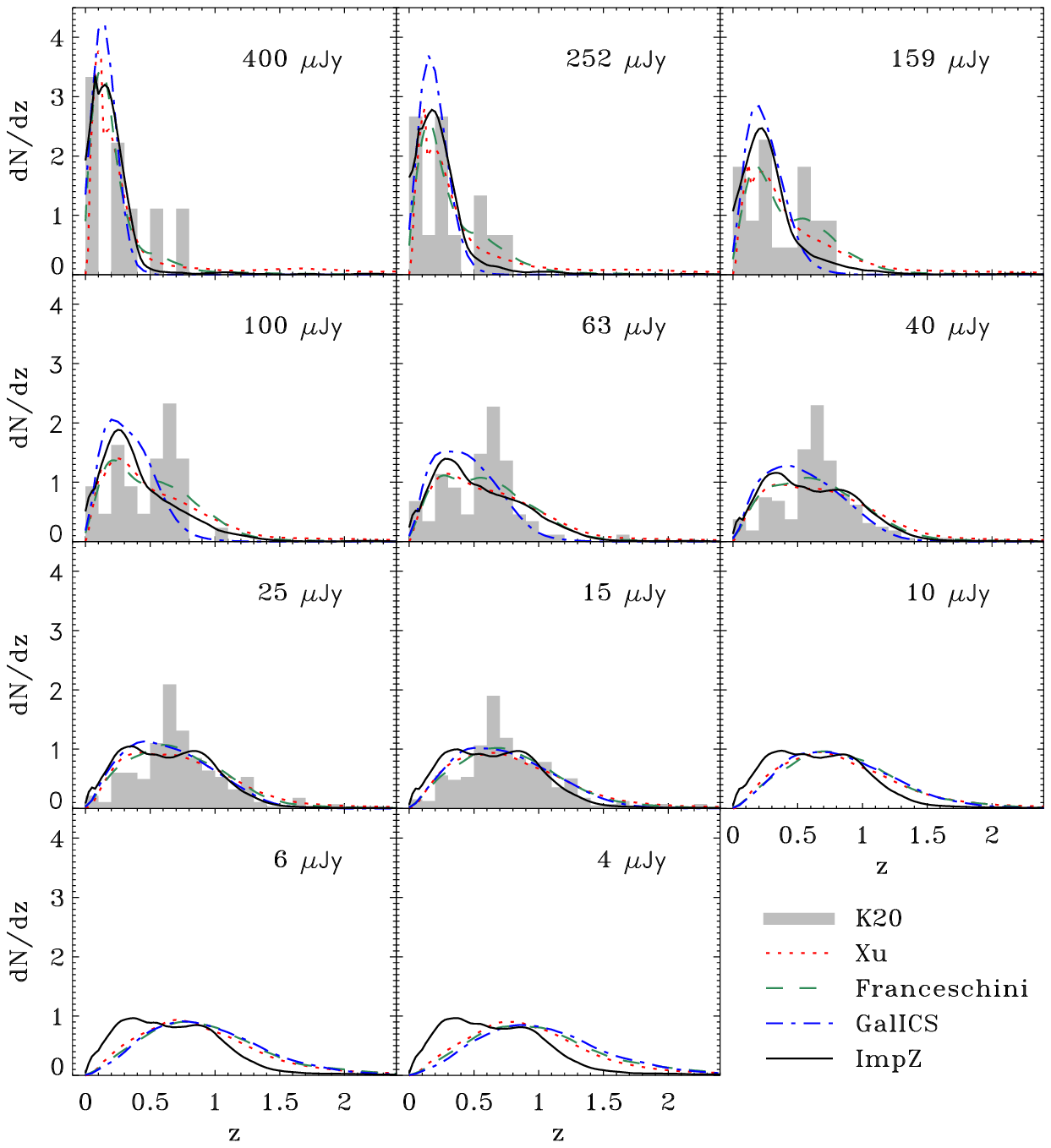}
\caption{The redshift distributions for each of the SWIRE flux-limited
  samples derived from the phenomenological models of \citet{Xu03} and
  \citet{Franceschini06}, the GalICS simulations \citep{Hatton03}, and
  photometric redshifts from {\sc ImpZ} \citep{Babbedge04}.  For
  comparison, the spectroscopic redshift distribution from the K20
  survey \citep{Mignoli05} is shown in the grey histograms.  Each
  distribution has been normalized over $0\le z< 5$.}
\label{fig:dndz}
\end{minipage}
\end{figure*}

The first two distributions were those predicted from the
phenomenological models of \citet{Xu03} and \citet{Franceschini06}.
These models combine local luminosity functions with parametric
modeling of luminosity and/or density evolution to predict the
relative numbers of different galaxy populations as a function of
redshift.  Several populations are defined (for example, early-type,
late-type, starburst, AGN), with spectral energy distributions drawn
from observed or model template libraries, giving multi-wavelength
predictions for the evolution of the galaxy populations.  These models
fit a wide range of observational data, in particular the mid-infrared
number counts from {\it Spitzer\/} surveys.

The GalICS simulations \citep{Hatton03} provided the third redshift
distribution, and the fourth estimate of $dN/dz$ was based on the
SWIRE survey directly, using the {\sc ImpZ} photometric redshift
catalogue \citep{Babbedge04,Rowan-Robinson05}.  The redshift
distributions are shown in Fig.~\ref{fig:dndz} and their median
redshifts are listed in Table~\ref{tab:amplitudes2}.  The median
redshifts are in good agreement with each other for flux limits of
$S_{36}= 25$~\mujy\ and brighter, corresponding to median redshifts of
$\left<z\right><0.7$--0.8.  Fainter than this, the model distributions
continue to shift to higher redshift with decreasing flux limit, but
the median redshift of the observational estimate ({\sc ImpZ}) remains
constant.  This primarily reflects the incompleteness of the optical
identifications of the SWIRE survey, as the photometric redshift code
is driven by the optical data.  We find that only 40~per cent of the
4~\mujy\ sample have photometric redshifts compared with $>90$~per
cent of sources brighter than 40~\mujy.  The sources without redshifts
are those that are optically faint and so are more likely to be
galaxies at high redshift.

Also plotted in Fig.~\ref{fig:dndz} (shaded histograms) are the
redshift distributions of the K20 survey \citep{Cimatti02,Mignoli05}.
The K20 survey is a near-infrared ($K_s<20$, Vega system) redshift
survey of 545 sources, with a high spectroscopic completeness of 92
per cent.  Using the equivalent $K$-band flux limits of the SWIRE
samples (section~\ref{sec:kband}), we compared our estimates of the
SWIRE redshift distributions with the observed K20 data.  The
spectroscopic redshift distributions are consistent with the SWIRE
estimates, however they do not distinguish between the different
models or the {\sc ImpZ} photometric redshifts -- there are simply too
few sources with spectroscopic redshifts, particularly in the brighter
samples.  One can also see a peak in the K20 distributions at
$z\simeq0.7$, corresponding to a cluster or other large-scale
structure in the K20 survey.  

This illustrates how the small size of current $K$-band selected
redshift surveys (typically less than 1000 sources) restricts their
usefulness in defining a redshift distribution, due to small-number
statistics and cosmic variance.  This, together with significant
selection biases (typically they target high-redshift or very red
sources), makes them unsuitable as redshift distributions for
calculating the inversion of Limber's equation, hence we use the
photometric and model estimates in the following sections.

\subsection{Spatial correlation lengths}

\begin{figure}
\includegraphics[width=\hsize]{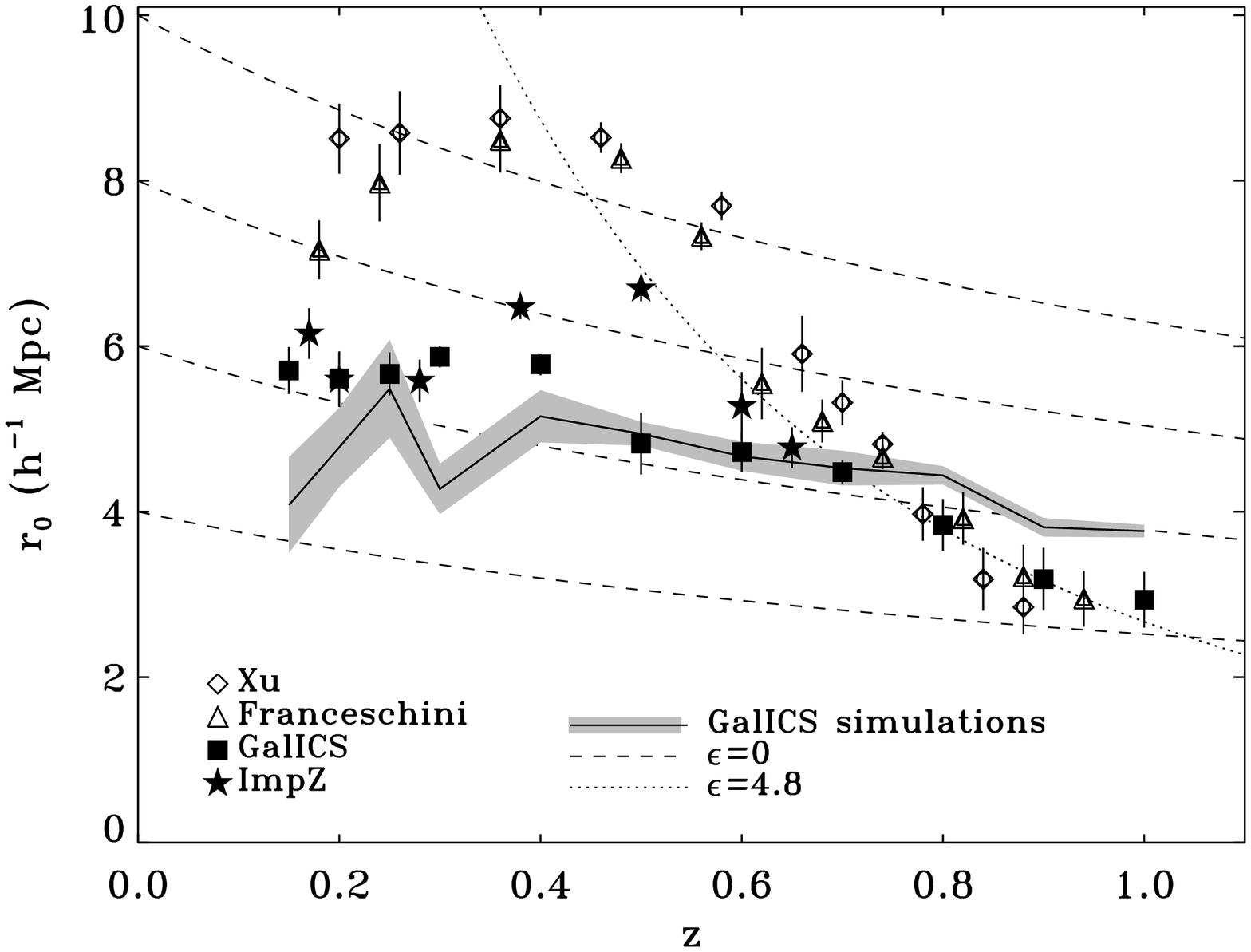}
\caption{The comoving correlation length, $r_0$, calculated from the
  SWIRE clustering amplitudes, as a function of median redshift,
  derived from each of the redshift distributions: \citet{Xu03},
  \citet{Franceschini06}, GalICS \citep{Hatton03} and {\sc ImpZ}
  \citep{Babbedge04}.  The solid line shows the results derived from
  the angular clustering measured from the GalICS simulations, with
  one-sigma error bounds shown in grey.  The dashed lines are the
  stable clustering ($\epsilon=0$) parametric models for different
  normalizations at $z=0$.  The dotted line is the best-fitting
  epsilon model at $z>0.45$, with $\epsilon=4.8\pm0.8$ and
  $r_0=27\pm6h^{-1}$~Mpc.}
\label{fig:r0zmulti}
\end{figure}

Taking each of the redshift distributions in turn, we have used the
inverse of Limber's equation (Eq.~\ref{eq:limber}) to calculate the
correlation length, $r_0$, from the angular clustering amplitude of
each of the samples.  Setting $\epsilon=\gamma-3=-1.2$ in
Eq.~\ref{eq:xi} for the case of comoving clustering gives a comoving
value of $r_0$.  The correlation lengths are given in
Table~\ref{tab:amplitudes2} and are plotted as a function of median
redshift in Fig.~\ref{fig:r0zmulti} for each $dN/dz$ (the plot excludes
the {\sc ImpZ} results for the faintest four samples, which are
incomplete).



For the five brightest samples, corresponding to
$\left<z\right>\la0.5$, the correlation length varies slowly
with redshift for each of the $dN/dz$ distributions, but with
an apparent dichotomy between the phenomological models with
$r_0\simeq8h^{-1}$~Mpc and the {\sc ImpZ} and GalICS estimates with
$r_0\simeq6h^{-1}$~Mpc.  The uncertainty in the correlation length is
dominated by this scatter between the $dN/dz$ estimates, not by the
statistical uncertainty in measuring the angular clustering.

For all $dN/dz$ distributions, the correlation length decreases
rapidly with increasing redshift in the fainter samples
(Fig.~\ref{fig:r0zmulti}), falling from $r_0\simeq6$--$8h^{-1}$~Mpc at
$z\simeq0.5$ to $3h^{-1}$~Mpc at $z\simeq1$.  Although the scatter in
the median redshifts of the distributions are larger in these fainter
samples, the average correlation lengths are more tightly constrained
than than those at lower redshift.  (This is easily understood: the
scatter between the median values of each redshift distribution is
small compared with the actual widths of the distributions, but at
lower redshifts this is not the case.)  If this high-redshift
($z>0.45$) evolution is fitted by an epsilon model (Eq.~\ref{eq:xi}),
we get best-fitting values of $r_0=27\pm6h^{-1}$~Mpc and
$\epsilon=4.8\pm0.8$ (Fig.~\ref{fig:r0zmulti}, dotted line).  This is
very strong evolution, and implies that these sources are in
environments that would evolve into massive clusters by $z=0$ (c.f.\
Fig.~\ref{fig:r0z}).  Although we do not consider the epsilon model to
be a realistic model of evolution to the present day, it does give
some quantitive indication of the rapid change in clustering seen at
these redshifts.

The spatial correlation lengths for the GalICS simulated catalogues
have been calculated in exactly the same way as those for the SWIRE
observations.  The amplitudes of the angular clustering
(section~\ref{sec:galics}) plus the redshift distributions
(section~\ref{sec:zdists}) of the GalICS samples were used to
calculate $r_0$ through the inverse of Limber's equation
(Eq.~\ref{eq:limber}).  These correlation lengths are also plotted in
Fig.~\ref{fig:r0zmulti} (solid line, with the one-sigma errors in
grey) for comparison with the results inferred from the observations.
The GalICS simulations closely follow a stable clustering model, with
a present-day correlation length of $r_0\simeq 6h^{-1}$~Mpc.  As
expected from the good agreement between the angular correlation
functions of the SWIRE data and the GalICS simulations, the spatial
correlation lengths from GalICS (grey region in
Fig.~\ref{fig:r0zmulti}) are consistent with those of SWIRE based on
the GalICS redshift distributions (squares).

In Fig.~\ref{fig:r0zmulti}, we see that the correlation lengths
derived from the GalICS $dN/dz$ (both the SWIRE values and the
simulations) differ markedly from the results based on the models of
\citet{Franceschini06} and \citet{Xu03}.  The latter predict that the
SWIRE results are due to stronger clustering at low median redshifts,
with rapid evolution at $z\ga0.5$, which is in contrast to the slowly
evolving stable clustering derived from the GalICS simulations.  The
results based on the {\sc ImpZ} photometric redshifts generally follow
the GalICS data but are noticably higher at $z\simeq0.5$, suggesting
more complex evolution than either the models or simulations predict.
Given that the {\sc ImpZ} redshift distributions are based on
empirical data rather than models, we consider these results to be the
best estimate of the true correlation lengths, adopting the GalICS
results for samples fainter than $S_{36}=25$~\mujy\ where the {\sc
ImpZ} data are incomplete.  These correlation lengths are reproduced
in Table~\ref{tab:amplitudes} alongside the angular measurements.


\begin{table}
\caption{Average 3.6~\micron\ luminosities, 
  $\log_{10}(L_{3.6}/L_\odot)$, and redshifts, 
  $\left<z\right>$, for the \citet{Xu03} and 
  \citet{Franceschini06} model redshift 
  distributions.  $S_{36}$ are the flux limits (\mujy). }
\begin{tabular}{ccccc} \hline
 &   \multicolumn{2}{c}{Xu et al.}           &   \multicolumn{2}{c}{Franceschini et al.} \\
$S_{36}$ & $\left<z\right>$ & $\log_{10}(L_{3.6}/L_\odot)$ 
         & $\left<z\right>$ & $\log_{10}(L_{3.6}/L_\odot)$ \\ 
  \hline
  4.0&0.88&$ 9.45\pm0.48$&0.94&$ 9.24\pm0.55$\\
  6.3&0.84&$ 9.50\pm0.45$&0.88&$ 9.29\pm0.52$\\
 10.0&0.78&$ 9.57\pm0.42$&0.82&$ 9.39\pm0.50$\\
 15.9&0.74&$ 9.64\pm0.40$&0.74&$ 9.44\pm0.48$\\
 25.2&0.70&$ 9.72\pm0.39$&0.68&$ 9.54\pm0.45$\\
 40.0&0.66&$ 9.80\pm0.39$&0.62&$ 9.64\pm0.42$\\
 63.4&0.58&$ 9.87\pm0.40$&0.56&$ 9.64\pm0.43$\\
100.5&0.46&$ 9.91\pm0.42$&0.48&$ 9.74\pm0.48$\\
159.2&0.36&$ 9.91\pm0.45$&0.36&$ 9.84\pm0.47$\\
252.4&0.26&$ 9.89\pm0.46$&0.24&$ 9.79\pm0.48$\\
400.0&0.20&$ 9.84\pm0.43$&0.18&$ 9.64\pm0.50$\\
  \hline
\end{tabular}
\label{tab:lum}
\end{table}

\begin{figure*}
\begin{minipage}{180mm}
\includegraphics[width=\hsize]{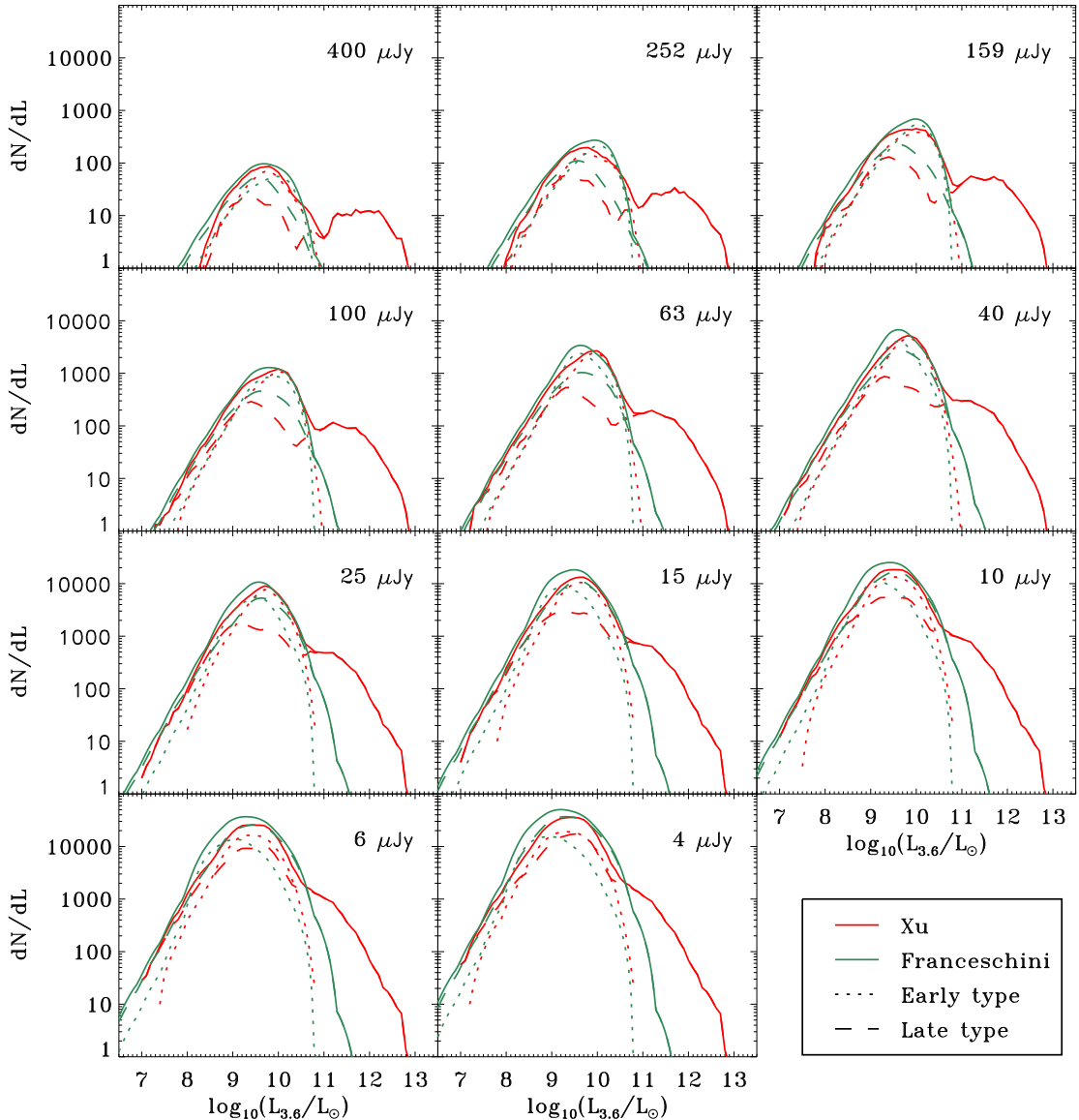}
\caption{The distribution of 3.6~\micron\ luminosities for each of the
eleven flux-limited samples from the \citet[][green]{Franceschini06}
and \citet[][red]{Xu03} models.  $dN/dL$ is the number of sources
per square degee per $\log_{10}L_{3.6}$.  Dotted lines are the early
types, dashed lines are the late types and solid lines are the total
number of sources.}
\label{fig:lumdist}
\end{minipage}
\end{figure*}

So far we have interpreted the change in $r_0$ as evolution in the
clustering strength, but another possibility is that we are looking at
different populations of sources in the different samples.  For
example, the fainter samples may be dominated by less massive galaxies
which are located in halos of lower mass and so are intrinsically less
clustered \citep[e.g.,][]{Loveday95,Norberg02,Zehavi05}.  We have
investigated this possibility using the 3.6-\micron\ luminosity
($L_{3.6}$) as a proxy for stellar mass.  In Fig.~\ref{fig:lumdist}
we plot the luminosity distributions of each sample from the
\citet{Franceschini06} and \citet{Xu03} models, showing the
contribution from early-type (elliptical/lenticular) and late-type
(spiral, irregular, starburst) populations separately.  For
3.6-\micron\ luminosities below $10^{11}$~L$_\odot$, the two models
predict very similar distributions for the total counts, but at higher
luminosities the \citet{Xu03} models have an extended tail of luminous
late-type galaxies that are not present in the \citet{Franceschini06}
models.  At bright flux limits ($S_{36}\ge40$~\mujy), the late-types
are about 30--40 per cent of the total number of galaxies.  The
fraction of late-types then increases with decreasing flux limit,
until in the faintest sample the late-types are 50~per cent of the
total according to the \citet{Xu03} model and 75~per cent of the total
in the \citet{Franceschini06} model.  In all cases, the redshift
distributions of the early-type galaxies are weighted towards higher
redshifts than those of the late-type galaxies.

\begin{figure}
\includegraphics[width=\hsize]{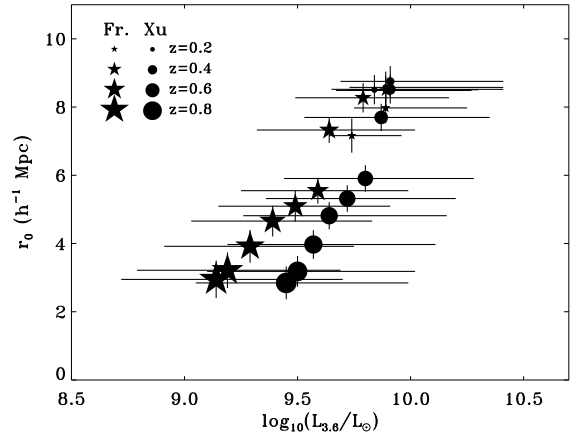}
\caption{Correlation lengths, $r_0$, as a function of 3.6~\micron\
luminosity, $\log_{10}(L_{3.6}/L_\odot)$, derived from the
\citet[][stars]{Franceschini06} and \citet[][circles]{Xu03} model
redshift distributions.  The symbol size is proportional to the median
redshift of each sample.  Luminosity errors are the standard
deviations of the luminosity distributions.}
\label{fig:r0lum}
\end{figure}

\begin{figure}
\includegraphics[width=\hsize]{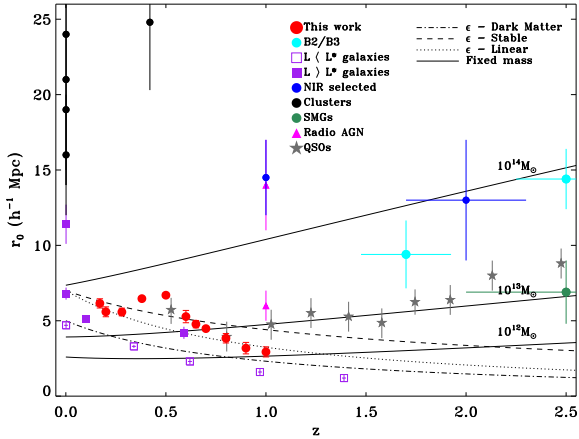}
\caption{The best estimate of the comoving correlation length, $r_0$, of our
samples (red circles) as a function of redshift, compared with values
from the literature \citep[][and references therein]{Farrah06}.  The
solid lines are the models of \citet{Matarrese97} for halos of fixed
mass, with masses as indicated.  Also shown are indicative $\epsilon$
models of clustering evolution: stable and linear models with
$r_0=7h^{-1}$~Mpc, corresponding to the average of our low-$z$
results; and a linear model normalized to the dark matter clustering
strength, $r_0=5h^{-1}$~Mpc, of \citet{Jenkins98}.}
\label{fig:r0z}
\end{figure}

In Fig.~\ref{fig:r0lum} we plot the spatial correlation lengths
against the average 3.6-\micron\ luminosities for the two models, and
tabulate the luminosities in Tab.~\ref{tab:lum}.  For the brightest
samples, i.e.\ the lowest median redshifts, the average luminosity is
approximately constant, at $\log_{10}(L_{3.6}/L_\odot)\simeq 9.8$.  At
fainter flux limits ($S_{36}=40$~\mujy\ and below), it is seen that
there is a trend for $r_0$ to decrease with decreasing average
luminosity.  Recalling that there is a progressively larger fraction
of late-type galaxies as the flux limit decreases, both the decrease
in average luminosity (mass) and the weaker clustering could be a
consequence of the changing galaxy population -- i.e.\ an increase in
the number of lower-mass late-type galaxies.  However, with the data
we have available we cannot unambiguously disentangle evolution in the
clustering strength from evolution in the relative populations of
early-types and late-types.

Finally, we have compared the best estimates of the correlation
lengths (Table~\ref{tab:amplitudes}) with values of $r_0$ from the
literature.  This is shown in Fig.~\ref{fig:r0z} where we plot $r_0$
as a function of redshift for our data, together with a variety of
galaxy and AGN correlation lengths, and several models, compiled by
\citet{Farrah06}.  The SWIRE sources are the most clustered population
of galaxies at $z<1$, with only clusters of galaxies exceeding the
correlation lengths of the SWIRE samples.  The solid lines are the
halo models of \citet{Matarrese97} which allow us to estimate the
approximate masses of the halos that host the 3.6-\micron\ selected
sources.  At $z\la0.5$ the inferred halo masses are approximately
constant with redshift with $\log(\rm{M_{Halo}/M_\odot})\sim 13.5$.
The correlation lengths have been shown above to decrease with
increasing redshift beyond $z=0.5$ and this corresponds to a decrease
in halo mass from $10^{13.5}$~M$_\odot$ to $10^{12}$~M$_\odot$ at
$z=1$.  This is consistent with the idea that we are detecting more
late-type galaxies, which have lower masses than the early-types that
are prevalent in the brighter samples.

\section{Conclusions}\label{sec:conclude}

The SWIRE survey has allowed us to measure the angular clustering of
sources selected at 3.6~\micron\ over larger scales and with greater
significance than previously has been possible.  The two-point angular
correlation function has been measured in eleven flux-limited samples
down to $S_{36}=4$~\mujy, corresponding to median redshifts $z\le1$.
These angular results are in good agreement with the GalICS
simulations and with $K$-band clustering measurements.

We have used Limber's equation to infer the spatial correlation
lengths from the angular measurements and four estimates of the
redshift distributions of each sample.  We find that the systematic
uncertainty in the $dN/dz$ distribution dominates the statistical
errors, and adopt the photometric redshift distribution from {\sc
ImpZ} as the best estimate of $dN/dz$, extrapolating to higher
redshifts using the GalICS results.  The comoving correlation length
varies slowly around $r_0=6.1\pm0.5h^{-1}$~Mpc out to a median
redshift of $z=0.5$ and then decreases with increasing redshift.  If
this change at higher redshifts is due to evolution in the clustering,
then the required evolution is very strong (with
$r_0=27\pm6h^{-1}$~Mpc and $\epsilon=4.8\pm0.8$) and these SWIRE
galaxies would be in environments that will evolve into massive
clusters by the present day.  There is some indication from the
\citet{Franceschini06} and \citet{Xu03} phenomological models that the
change in correlation length is due to a decrease in the average
3.6-\micron\ luminosity (or equivalently, stellar mass) of the fainter
samples rather than evolution in the clustering.  This decrease in
average luminosity is the result of an increased fraction of late-type
galaxies in the faint samples.  Comparing these SWIRE data with the
halo models of \citet{Matarrese97} suggests that the fainter samples
are selecting sources in lower mass halos.  These latter two results
provide a consistent picture where lower-mass late-type galaxies are
preferentially found in lower mass halos and thus have weaker
clustering.

The comparison of our data with the GalICS simulations has shown that
these models are a good match to the angular clustering, but their
redshift distributions differ markedly from the phenomological models
and the photometric redshift distributions.  This highlights the need
for a better understanding of the spatial clustering and its
evolution.  Higher-resolution numerical models and improved
semi-analytic models will allow us to directly explore the relation
between the clustering of the galaxies and the dark matter, i.e.\ the
bias.  For example, the dark matter Millenium Simulation
\citep{Springel05} has a lower mass resolution limit than GalICS, and
combined with semi-analytic models of the galaxies it is a better
match to the area and depth of SWIRE than the current models.  Future
work will explore the comparison between these models and the SWIRE
data presented here.

\section*{Acknowledgements}

The pair-counting code {\sc npt} was provided by The Auton Lab at
Carnegie Mellon University's School of Computer Science
(http://www.autonlab.org) and we thank Andrew Moore et al.\ for making
this software public.  This work was funded by PPARC research grants
to Seb Oliver.  Support for this work, part of the {\it Spitzer\/}
Space Telescope Legacy Science Programme, was provided by NASA through
a contract issued by the Jet Propulsion Laboratory, California
Institute of Technology under a contract with NASA.

\bsp

\label{lastpage}


\begin{thebibliography}{}

\bibitem[Babbedge et al.(2004)]{Babbedge04} Babbedge, T.~S.~R., 2004,
  \mnras, 353, 654

\bibitem[Bardeen et al.(1986)]{Bardeen86} Bardeen, J.~M., Bond, J.~R.,
  Kaiser, N., Szalay, A.~S. 1986, \apj, 304, 15

\bibitem[Baugh et al.(1996)]{Baugh96} Baugh, C.~M., Gardner, J.~P.,
  Frenk, C.~S., Sharples, R.~M . 1996, \mnras, 283, L15

\bibitem[Benson et al.(2001)]{Benson01} Benson, A. J., Frenk, C. S.,
  Baugh, C. M., Cole, S., Lacey, C. G. 2001, \mnras, 327, 1041

\bibitem[Berlind \& Weinberg(2002)]{Berlind02} Berlind, A.~A.,
  Weinberg, D.~H. 2002, \apj, 575, 587

\bibitem[Bertin \& Arnouts(1996)]{Bertin96} Bertin, E., Arnouts, S.,
  1996, \aass, 117, 393

\bibitem[Blaizot et al.(2006)]{Blaizot06} Blaizot, J., et al. 2006,
  \mnras, 369, 1009

\bibitem[Cimatti et al.(2002)]{Cimatti02} Cimatti, A., et al. 2002,
  \aas, 392, 395

\bibitem[Cole et al.(2000)]{Cole00} Cole, S., Lacey, C.~G., Baugh,
  C.~M., Frenk, C.~S. 2000, \mnras, 319, 168

\bibitem[Dye et al.(2006)]{Dye06} Dye, S., et al., 2006, \mnras,
  submitted, astro-ph/0603608

\bibitem[Efstathiou et al.(1991)]{Efstathiou91} Efstathiou, G.,
  Bernstein, G., Katz, N., Tyson, J. A., Guhathakurta, P. 1991, \apjl,
  380, L47

\bibitem[Fang et al.(2004)]{Fang04} Fang, F., et al. 2004, \apjs, 154,
  35

\bibitem[Farrah et al.(2006)]{Farrah06} Farrah, D., et al. 2006,
  \apjl, 641, L17, and Erratum, 2006, \apjl, 643, L139

\bibitem[Fazio et al.(2004)]{Fazio04} Fazio, G. G., et al. 2004,
  \apjs, 154, 39

\bibitem[Franceschini et al.(2006)]{Franceschini06} Franceschini, A.,
  et al. 2006, \aas, 453, 397

\bibitem[Gonzalez-Solares et al.(2004a)]{Gonzalez-Solares04}
  Gonzalez-Solares, E., et al. 2004a, \mnras, 352, 44 

\bibitem[Gonzalez-Solares et al.(2004b)]{Gonzalez-Solares04b}
  Gonzalez-Solares, E., et al. 2004b, \mnras, 358, 333 

\bibitem[Granato et al.(2000)]{Granato00} Granato, G.~L., Lacey, C.~G.,
  Silva, L., Bressan, A., Baugh, C.~M., Cole, S., Frenk, C.~S.  2000,
  \apj, 542, 710

\bibitem[Gray et al.(2004)]{Gray04} Gray, A. G., Moore, A. W., Nichol,
  R. C., Connolly, A. J., Genovese, C., Wasserman, L. 2004, in
  F. Ochsenbein, M. Allen and D. Egret (eds), Astronomical Data Analysis
  Software and Systems XIII, ASP Conf Series, 314, 249

\bibitem[Hatton et al.(2003)]{Hatton03} Hatton, S., Devriendt,
  J.~E.~G., Ninin, S., Bouchet, F.~R., Guiderdoni, B., \& Vibert, D.\
  2003, \mnras, 343, 75

\bibitem[Jenkins et al.(1998)]{Jenkins98} Jenkins, A., et al. 1998,
  \apj, 499, 20 

\bibitem[Kaiser(1984)]{Kaiser84} Kaiser, N. 1984, \apjl, 284, L9

\bibitem[Kong et al.(2006)]{Kong06} Kong, X., et al. 2006, \apj, 638,
  72

\bibitem[K\"{u}mmel \& Wagner(2000)]{Kummel00} K\"{u}mmel, M.~W., 
  Wagner, S.~J. 2000, \aas, 353, 867

\bibitem[Landy \& Szalay(1993)]{Landy93} Landy, S. D., Szalay,
  A. S. 1993, \apj, 412, 64

\bibitem[Limber(1953)]{Limber53} Limber, D.~N., 1953, \apj, 117, 
134 

\bibitem[Lonsdale et al.(2003)]{Lonsdale03} Lonsdale, C. J., et
  al. 2003, \pasp, 115, 897

\bibitem[Lonsdale et al.(2004)]{Lonsdale04} Lonsdale, C. J., et
  al. 2004, \apjs, 154, 54

\bibitem[Loveday et al.(1995)]{Loveday95} Loveday, J., Maddox, S. J.,
  Efstathiou, G., Peterson, B. A. 1995, \apjl, 442, 457L

\bibitem[Matarrese et al.(1997)]{Matarrese97} Matarrese, S., Coles,
  P., Lucchin, F., Moscardini, L. 1997, \mnras, 286, 115

\bibitem[McCrcaken et al.(2000)]{McCracken00} McCracken, H., Shanks, T.,
  Metcalfe, N., Fong, R., Campos, A. 2000, \mnras, 318, 913

\bibitem[McMahon et al.(2001)]{McMahon01} McMahon, R. G., Walton,
  N. A., Irwin, M. J., Lewis, J. R., Bunclark, P. S., Jones, D. H.,
  2001, New Astron. Rev., 45, 97

\bibitem[Mignoli et al.(2005)]{Mignoli05} Mignoli, M., et al. 2005,
  \aas, 437, 883

\bibitem[Moscardini et al.(1998)]{Moscardini98} Moscardini, L., Coles,
  P., Lucchin, F., Matarrese, S. 1998, \mnras, 299, 95

\bibitem[Norberg et al.(2002)]{Norberg02} Norberg, P., et al. 2002,
  \mnras, 332, 827

\bibitem[Oliver et al.(2000)]{Oliver00} Oliver, S. J., et al. 2000,
  \mnras, 316, 749

\bibitem[Oliver et al.(2004)]{Oliver04} Oliver, S. J., et al. 2004,
  \apjs, 154, 30

\bibitem[Overzier et al.(2003)]{Overzier03} Overzier, R. A.,
  R\"ottgering, H. J. A., Rengelink, R. B., Wilman, R. J., 2003, \aas,
  405, 53

\bibitem[Phillipps et al.(1978)]{Phillipps78} Phillipps, S., Fong,
  R., Ellis, R. S., Fall, S. M., MacGillivray, H. T., 1978, \mnras,
  182, 673

\bibitem[Pollo et al.(2005)]{Pollo05} Pollo, A., et al. 2005, \aas,
  439, 887

\bibitem[Reike et al.(2004)]{Reike04} Reike, G., et al. 2004, \apjs,
  154, 25

\bibitem[Roche et al.(1998)]{Roche98} Roche, N., Eales, S.,
  Hippelein, H., 1998, \mnras, 295, 946

\bibitem[Roche et al.(1999)]{Roche99} Roche, N., Eales, S. A.,
  Hippelein, H., Willott, C. J. 1999, \mnras, 306, 538

\bibitem[Roche et al.(2002)]{Roche02} Roche, N., Almaini, O., Dunlop,
  J., Ivison, R.~J., Willott, C.~J. 2002, \mnras, 337, 1282

\bibitem[Roche et al.(2003)]{Roche03} Roche, N., Dunlop, J. S.,
  Almaini, O. 2003, \mnras, 346, 803

\bibitem[Rowan-Robinson et al.(2004)]{Rowan-Robinson04}
  Rowan-Robinson, M., et al. 2004, \mnras, 351, 1290

\bibitem[Rowan-Robinson et al.(2005)]{Rowan-Robinson05}
  Rowan-Robinson, M., et al. 2005, \aj, 129, 1183

\bibitem[Scranton et al.(2002)]{Scranton02} Scranton, R., et al. 2002,
  \apj, 579, 48

\bibitem[Skrutskie et al.(2006)]{Skrutskie06} Skrutskie, M. F., et
  al. 2006, \aj, 131, 1163

\bibitem[Springel et al.(2005)]{Springel05} Springel, V., et al. 2005,
  Nature, 435, 629

\bibitem[Surace et al.(2005)]{Surace05} Surace, J. A., et al. 2005,
  `The SWIRE Data Release 2: Image Atlases and Source Catalogs for
  ELAIS-N1, ELAIS-N2, XMM-LSS and the Lockman Hole', Spitzer Science
  Center, California Institute of Technology (Pasadena, CA)

\bibitem[Surace et al.(2006)]{Surace06} Surace, J. A., et al. 2006,
  \apjs, in preparation

\bibitem[Werner et al.(2004)]{Werner04} Werner, M., et al. 2004, \apjs,
  154, 1

\bibitem[Xu et al.(2003)]{Xu03} Xu, C. K., Lonsdale, C. J., Shupe,
  D. L., Franceschini, A., Martin, C., Schiminovich, D. 2003, \apj,
  587, 90

\bibitem[York et al.(2000)]{York00} York, D. G., et al. 2000, \aj,
  120, 1579

\bibitem[Zehavi et al.(2002)]{Zehavi02} Zehavi, I., et al. 2002, \apj,
  571, 172

\bibitem[Zehavi et al.(2005)]{Zehavi05} Zehavi, I., et al. 2005, \apj,
  621, 22



\end{thebibliography}
\end{document}